\documentclass[reprint,aps,prl,showpacs,showkeys]{revtex4-1}

\usepackage{natbib}
\usepackage{graphics}
\usepackage{bm}
\usepackage{hyperref}

\begin{document}

\title{Electromagnetic Oscillations in a Driven Nonlinear Resonator:\\
A New Description of Complex Nonlinear Dynamics}
\author{E.~Yu.~Petrov}
\author{A.~V.~Kudrin}
\email{kud@rf.unn.ru}
\affiliation{Department of Radiophysics, University of Nizhny Novgorod,
23 Gagarin Ave., Nizhny Novgorod 603950, Russia}
\begin{abstract}
Many intriguing
properties of driven nonlinear resonators,
including the appearance of chaos,
are very important for understanding the universal
features of nonlinear dynamical systems and can have
great practical significance.
We consider a cylindrical cavity
resonator driven by an alternating voltage and filled
with a nonlinear nondispersive medium. It is assumed that
the medium lacks a center of inversion and the dependence
of the electric displacement on the electric
field can be approximated by an exponential function. We show
that the Maxwell equations are integrated
exactly in this case and the field components in the cavity are
represented in terms of implicit functions of special
form.
The driven electromagnetic oscillations in the cavity are found to
display very
interesting temporal behavior and their Fourier spectra contain singular
continuous components. To the best of our knowledge, this is the first
demonstration of the existence of a singular continuous (fractal)
spectrum in an exactly integrable system.
\end{abstract}


\maketitle

Nonlinear resonators are simple and convenient models of
physical systems and have proven to be a valuable means
to investigate the universal features of nonlinear dynamics.
In view of this, such resonators have been the subject of
intense theoretical and experimental studies in the past
decades~\cite{Lin,Bus,Bax,Iva}. A variety of the existing electronic
devices and materials with nonlinear electromagnetic
properties makes it possible to create electrical resonators
with different types of nonlinearity.
From a practical viewpoint, it is important that
when  electromagnetic oscillations  in a bounded volume are excited at
frequencies close to resonant ones, oscillation
amplitudes that are sufficient for manifestation of
various nonlinear phenomena can be reached even if a driving source is comparatively
weak. The well-known example of such phenomena is generation
of harmonics of a drive frequency, which enables
realization of efficient frequency multipliers on the basis of
high-Q resonators. In addition, fairly complex, e.g., chaotic, oscillations
can be excited in nonlinear resonators~\cite{Lin,Bus,Bax}. Studying chaotic
regimes of resonators is of great interest for developing controllable
sources of noise-like signals. In more recent works, a new type of complex
nonlinear dynamics, intermediate between almost periodic and
random, has been discovered~\cite{Gre,Aub,Luc}. Such  dynamics, which is associated
with a singular continuous spectrum, appears typically in
driven nonlinear systems and has received much
attention~\cite{Din,P94,P95,Yal,Bez,Agr,Zak}, both
for fundamental reasons and because of its interdisciplinary
relevance~\cite{Luc}.

A complete description of the complex dynamics of
distributed nonlinear systems is fairly difficult to achieve.
This is explained by an infinite number of degrees of freedom
and the presence of several controlling parameters in such systems.
Because of this, most theoretical works on the subject discuss
nonlinear resonators as lumped systems or merely as a collection
of coupled oscillators or modes. Within the framework of such
a simplified approach, the problem of oscillations in a nonlinear
resonator is reduced to solving a system of ODEs. Although such an
approach is justified in many cases, it is clear that electromagnetic
systems should generally be described by the Maxwell equations.

In this work, the problem of a nonlinear electrical resonator
is considered using a full set of the Maxwell equations.
In what follows, we apply the method for constructing exact
axisymmetric solutions
of the Maxwell equations in a nonlinear nondispersive medium,
which has been developed in our recent works~\cite{Pet,Esk}, to the driven oscillations
in a bounded volume.

Consider electromagnetic fields in a
cylindrical cavity of radius $a$ and height $L$. We assume that
the $z$ axis of a cylindrical coordinate system ($r$, $\phi$, $z$)
is aligned with the cavity axis and limit ourselves to consideration
of axisymmetric field oscillations
in which only the $E_z$ and $H_\phi$ components are nonzero.
We will also assume that the cavity is filled with a nonlinear
nondispersive medium in which the longitudinal component
of the electric displacement can be represented
as $D_{z}=D_{0}+\alpha^{-1}\epsilon_{0}\varepsilon_{1}[\exp(\alpha E_{z})-1]$,
where $D_0$, $\varepsilon_1$, and $\alpha$ are certain constants.
The possibility of using such a model of nonlinearity for media lacking
a center of inversion is
discussed in detail in~\cite{Pet}. As is shown in~\cite{Pet},
this model, with appropriately chosen $D_0$, $\varepsilon_1$, and $\alpha$, correctly describes dielectric properties of such media in the case of moderately small electric fields. Then the Maxwell
equations read
\begin{equation}
\partial_{r}H+r^{-1}H=\varepsilon(E)\,\partial_{t}E,\quad
\partial_{r}E=\mu_{0}\,\partial_{t}H,\label{eq1}
\end{equation}
where $E\equiv E_{z}(r,t)$, $H\equiv H_{\phi}(r,t)$, and
\begin{equation}
\varepsilon(E)\equiv dD_{z}/dE=\epsilon_{0}\varepsilon_{1}\exp(\alpha E).\label{eq2}
\end{equation}
The exact solution
of system~(1) can be written in implicit form as~\cite{Pet}
\begin{eqnarray}
&&\hspace{-5mm}
\tilde{E}=A^{-1}{\cal E}\left(\rho\,e^{\tilde{\alpha} \tilde{E}/2},
\tau+\tilde{\alpha}\rho \tilde{H}/2\right),\nonumber\\
&&\hspace{-5mm}
\tilde{H}=e^{\tilde{\alpha} \tilde{E}/2}
A^{-1}{\cal H}\left(\rho\,e^{\tilde{\alpha} \tilde{E}/2},
\tau+\tilde{\alpha} \rho \tilde{H}/2\right).\label{eq3}
\end{eqnarray}
Hereafter, $A$ is a constant
amplitude factor related to the field source, $\tilde{E}=E/A$, $\tilde{H}=Z_{0}H/(A\varepsilon^{1/2}_{1})$, $\rho=r/a$, $\tau=t(\epsilon_{0}\varepsilon_{1}\mu_{0})^{-1/2}/a$, and $\tilde{\alpha}=\alpha A$, where
$Z_{0}=(\mu_{0}/\epsilon_{0})^{1/2}$. The functions
${\cal E}$ and ${\cal H}$ describe the electromagnetic field in a linear
medium and satisfy the equations
\begin{equation}
\partial^{2}_{\rho}{\cal E}+\rho^{-1}\partial_{\rho}{\cal E}=\partial^{2}_{\tau}{\cal E},
\label{eq4}
\end{equation}
and $\partial_{\rho}{\cal E}=\partial_{\tau}{\cal H}$.

Let us state the following initial and boundary conditions for linear
wave equation~(4):
\begin{eqnarray}
&&\hspace{-5mm}
{\cal E}(\rho,0)=0,\quad \partial_{\tau}{\cal E}(\rho,0)=0,
\quad 0\le \rho < 1,\label{eq5}\\
&&\hspace{-5mm}
{\cal E}(1,\tau)=A\,(\sin \Omega_{1}\tau+0.5\,\sin \Omega_{2}\tau), \quad 0\le \tau < \infty,\label{eq6}
\end{eqnarray}
where $\Omega_{1,2}$ are normalized constant frequencies such that $\Omega_{1,2}=\lambda_{1,2} a(\epsilon_{0}\varepsilon_{1}\mu_{0})^{1/2}$
(i.e., $\Omega_{1,2} \tau=\lambda_{1,2}t$).
The frequencies $\Omega_1$ and $\Omega_2$ are related as
$\Omega_{1}=\sigma\Omega_{2}$, where $\sigma=(\sqrt{5}-1)/2$ is the golden mean.
The boundary value problem defined by Eqs.~(\ref{eq4})--(\ref{eq6}) describes
the driven electromagnetic oscillations in a cylindrical cavity
specified by the relations $\rho=r/a\le 1$ and $0\le z \le L$,
which is filled with a linear medium having the permittivity
$\varepsilon=\epsilon_{0}\varepsilon_{1}={\rm const}$ ($\alpha=0$).
The linear oscillations are thus driven by an
electric field~(\ref{eq6}) which can be produced by two coaxial
metal rings of radius $a$ that are separated by distance $L$ if
an almost periodic voltage
$V={\cal E}(1,\tau)L=AL(\sin\lambda_{1}t+0.5\,\sin\lambda_{2}t)$ is applied
between them.
The solution
to the linear boundary value problem specified by Eqs.~(\ref{eq4})--(\ref{eq6})
can be found in a standard way~\cite{Cou}. As a result, the
functions ${\cal E}$ and ${\cal H}$ are written as
\begin{eqnarray}
{\cal E}(\rho,\tau)\!&=&\!\sum\limits_{j=1}^{2}\!B_{j}
\frac{J_{0}(\Omega_j\rho)}{J_{0}(\Omega_j)}
\sin\Omega_j\tau
\!+\!\sum\limits_{n=1}^{\infty}C_{n}J_{0}(\kappa_{n}\rho)\sin\kappa_{n}\tau,
\nonumber\\
{\cal H}(\rho,\tau)\!&=&\!\sum\limits_{j=1}^{2}\!B_{j}\frac{J_{1}(\Omega_{j}\rho)}
{J_{0}(\Omega_{j})}\cos\Omega_{j}\tau
\!+\!\sum\limits_{n=1}^{\infty}C_{n}J_{1}(\kappa_{n}\rho)\cos\kappa_{n}\tau,\nonumber\\
\label{eq7}
\end{eqnarray}
where $B_{j}=A/j$,  $C_{n}=2A\sum_{j=1}^{2}\Omega_{j}
[j(\Omega_{j}^{2}-\kappa^{2}_{n})J_{1}(\kappa_{n})]^{-1}$,
$J_{m}$ is the Bessel function of the first kind of order $m$, $\kappa_{n}$ is
the $n$th root of the equation $J_{0}(\kappa)=0$, and $\Omega_{1,2}\neq \kappa_{n}$.
We denote eigenfrequencies of the $E_{0n0}$ modes~\cite{Jac} of a linear resonator
as $\omega_{n}$. Hence, $\omega_{n}=\kappa_{n}(\epsilon_{0}\mu_{0})^{-1/2}a^{-1}$ and
$\kappa_{n}\tau=\omega_{n}t$.

Substituting functions~(\ref{eq7}) into formulas~(\ref{eq3}), we obtain an exact solution to system~(\ref{eq1})
in implicit form. Thus, the field components $E$ and $H$ in a cylindrical
cavity filled with a nonlinear medium for which $\varepsilon(E)$ is written
in the form of Eq.~(\ref{eq2}) are found from the solution of a set of transcendental equations~(\ref{eq3})
in which ${\cal E}$ and ${\cal H}$, defined by relationships (\ref{eq7}), are almost periodic functions of $\tau$.
In the nonlinear case, the fields $E$ and $H$, which are determined  by Eqs.~(\ref{eq3})
and~(\ref{eq7}),
satisfy the same initial conditions as in Eq.~(5):
\begin{eqnarray}
E(\rho,0)=0,\quad \partial_{\tau}E(\rho,0)=0,
\quad 0\le \rho < 1.\label{eq8}
\end{eqnarray}
However, the electric field oscillations on the side wall of the nonlinear
resonator ($\rho=1$) in the found solution do not obey Eq.~(6).
Putting $\rho=1$ in formulas~(3), we have
\begin{eqnarray}
&&\hspace{-5mm}
\tilde{E}=A^{-1}{\cal E}\left(e^{\tilde{\alpha} \tilde{E}/2},
\tau+\tilde{\alpha} \tilde{H}/2\right),\nonumber\\
&&\hspace{-5mm}
\tilde{H}=e^{\tilde{\alpha} \tilde{E}/2}
A^{-1}{\cal H}\left(e^{\tilde{\alpha} \tilde{E}/2},
\tau+\tilde{\alpha} \tilde{H}/2\right).\label{eq9}
\end{eqnarray}
The dependence $\tilde{E}(1,\tau)$ determined by relationships~(\ref{eq9}) can be
regarded as a drive signal in the boundary value problem given by Eqs.~(\ref{eq1}),
(\ref{eq8}), and~(\ref{eq9}) for a nonlinear ($\alpha\neq 0$) resonator.

Thus, formulas (\ref{eq3}), with ${\cal E}$ and ${\cal H}$ given
by relationships~(\ref{eq7}), yield
an exact solution of the nonlinear boundary value problem for
system~(\ref{eq1}) under conditions~(\ref{eq8}) and~(\ref{eq9}), and describe driven
electromagnetic oscillations in a cylindrical cavity resonator filled
with a nonlinear medium. A typical example of such a
medium can be a ferroelectric crystal.
Note that ferroelectric resonators are known to be
of great interest for many promising applications~\cite{Iva}.

Observe that the oscillations on the axis $\rho=0$ of the nonlinear
resonator in the obtained exact solution
coincide with the oscillations for $\rho=0$ in the ``seeding''
linear problem. It follows from Eqs.~(\ref{eq3}) and~(\ref{eq7}) that for $\rho=0$,
$E(0,\tau)\equiv {\cal E}(0,\tau)=\sum^{2}_{j=1}B_{j}\sin\Omega_{j}\tau+
\sum^{\infty}_{n=1}C_{n}\sin\kappa_{n}\tau$ and
$H(0,\tau)\equiv {\cal H}(0,\tau)=0$. Thus, the electric field
oscillations on the axis $\rho=0$ are described by an almost periodic function of $\tau$
and have the discrete spectrum~\cite{Boh}.

For $\rho\neq 0$, the exact solution expressed in terms of
implicit functions
is very complicated and can be studied only numerically.
It turns out that for $\rho\neq 0$, the field oscillations
described by this solution may have singular continuous (fractal)
spectrum.

To confirm the above assertion,
we turn to results of calculations of the quantities
$\tilde{E}$ and $\tilde{H}$ determined by Eqs.~(\ref{eq3}) and~(\ref{eq7}).
In what follows, the main attention
will be focused on analyzing the obtained solutions in the case where
the drive frequency $\lambda_2$
relates to the fundamental eigenfrequency $\omega_{1}$ of a linear resonator as
$\lambda_{2}/\omega_{1}=\Omega_{2}/\kappa_{1}=\sigma$
(note that the identity $\Omega_{1}+\Omega_{2}=\kappa_{1}$ is valid in this
case).
In our calculations, we retain 100 terms of the series over $n$ in formulas~(\ref{eq7}).
It should be noted
that the employed theoretical model of a nondispersive medium does not
allow one to indefinitely increase the nonlinearity parameter $\tilde{\alpha}$ in
solutions~(3). For large absolute values of $\tilde{\alpha}$, implicit functions
$\tilde{E}(\rho,\tau)$ and $\tilde{H}(\rho,\tau)$ determined by
Eqs.~(\ref{eq3}) and~(\ref{eq7}) become
ambiguous, and solutions~(\ref{eq3}) obtained without
allowance for dispersion will be inapplicable~\cite{Pet}. For fixed
$\tilde{\alpha}$ and
$\Omega_{1,2}$, the ambiguity points appear first in the time
dependences $\tilde{E}(1,\tau)$ and $\tilde{H}(1,\tau)$ (for $\rho=1$) since the
nonlinear effects are accumulating with increasing $\rho$~\cite{Pet}. Therefore,
as a first step in practical calculations, one should study the
functions $\tilde{E}(1,\tau)$ and $\tilde{H}(1,\tau)$. If these functions are unambiguous
and continuous, then $\tilde{E}$ and $\tilde{H}$ possess the same properties
for $0\le \rho<1$. In all the computations, we
use the maximum possible
value $\tilde{\alpha}=0.32$ for chosen $\Omega_{1,2}$.

Now consider the field oscillations on the wall $\rho=1$ of the
nonlinear resonator. The dependences $\tilde{E}(1,\tau)$
and $\tilde{H}(1,\tau)$ determined by relationships~(\ref{eq9}) are shown in
Fig.~1(a) by the red and blue
lines, respectively. For comparison, the solid and dashed black
lines in Fig.1~(a) show the functions ${\cal E}$ and ${\cal H}$, respectively.
It is seen in Fig.~1 that the functions $\tilde{E}$ and $\tilde{H}$ demonstrate fairly
complex behavior and essentially differ from the corresponding
quantities ${\cal E}/A$ and ${\cal H}/A$ in the linear regime
($\alpha=0$) by the presence
of small amplitude spikes.
Figures 1(b) and 1(c) show the frequency spectra of $\tilde{E}(1,\tau)$
and $\tilde{H}(1,\tau)$ ($2^{16}$ points were taken for the FFT with a
sampling rate of 0.05). The FFT plots exhibit numerous spectral components that
can be attributed to the harmonics of $\Omega_{1,2}$ and $\kappa_{n}$,
as well as various combination frequencies.

To reveal nontrivial spectral properties of the oscillations considered,
we use the singular-continuous spectrum analysis~\cite{Aub,P94,P95,Yal,Agr}.
We define the
partial Fourier sums
\begin{equation}
S_{E}(\Omega,T)=\sum\limits^{T}_{m=1}\tilde{E}_{m}e^{i\Omega\tau_{m}},\quad
S_{H}(\Omega,T)=\sum\limits^{T}_{m=1}\tilde{H}_{m}e^{i\Omega\tau_{m}},\label{eq10}
\end{equation}
where $\{\tilde{E}_{m}\}$ and $\{\tilde{H}_{m}\}$ are the time series of the variables
$\tilde{E}$ and $\tilde{H}$: $\tilde{E}_{m}=\tilde{E}(\tau_{m})$ and $\tilde{H}_{m}=\tilde{H}(\tau_{m})$.
We take $\tau_{1}=100$ and
$\tau_{m+1}-\tau_{m}=0.02$,
although all the forthcoming results are valid when changing the
initial point and the sampling rate.
The Fourier transforms scale with $T$ as
\begin{equation}
|S_{E}(\Omega,T)|^{2}\sim T^{\beta},\quad
|S_{H}(\Omega,T)|^{2}\sim T^{\gamma},\label{eq11}
\end{equation}
where $\beta=\beta(\Omega)$ and $\gamma=\gamma(\Omega)$ are scaling
exponents~\cite{Aub,P94,Yal}. The evolution of $S_{E}$ and $S_{H}$ with $T$ can be represented
by paths in the complex planes (${\rm Re}\,S_{E}$, ${\rm Im}\,S_{E}$) and
(${\rm Re}\,S_{H}$, ${\rm Im}\,S_{H}$), respectively.
It is known~\cite{P94} that for $\beta=\gamma=2$, the frequency $\Omega$ belongs
to the countable set of discrete spectral components of
an almost periodic oscillation and there exist persistent motions (drifts)
of $S_{E}$ and $S_{H}$ in the corresponding complex planes.
A singular-continuous spectral component appears if (i) $\beta\neq 1,2$ and/or
$\gamma\neq 1,2$ and (ii) the path in
the complex plane is fractal~\cite{Aub,P94,P95,Yal,Agr}.
A singular continuous spectrum is known to be a Cantor set~\cite{Luc,Zak}.

\begin{figure}[h]
\includegraphics{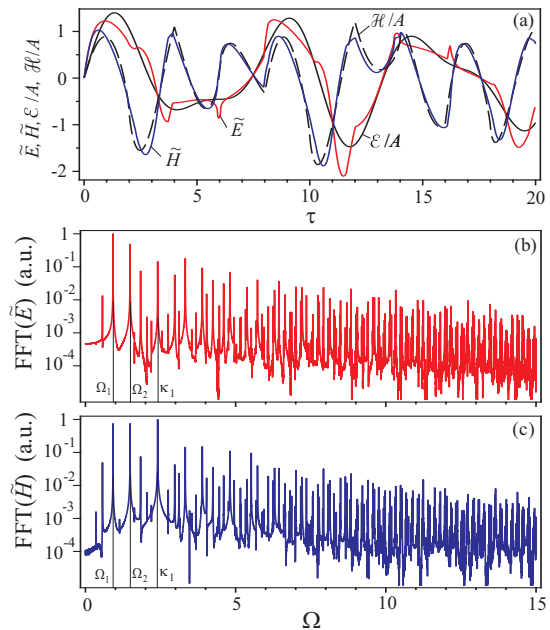}
\caption{(color online) (a)~Oscillograms of the electric ($\tilde{E}$) and magnetic
($\tilde{H}$) fields on the wall $\rho=1$ of a nonlinear resonator (red
and blue lines, respectively), calculated by formulas~(9), and the corresponding quantities
${\cal E}/A$ and ${\cal H}/A$ in the linear regime (solid and
dashed black lines, respectively). The fast Fourier
transforms  of  (b)~$\tilde{E}$ and (c)~$\tilde{H}$.}
\end{figure}

\begin{figure}[h]
\includegraphics{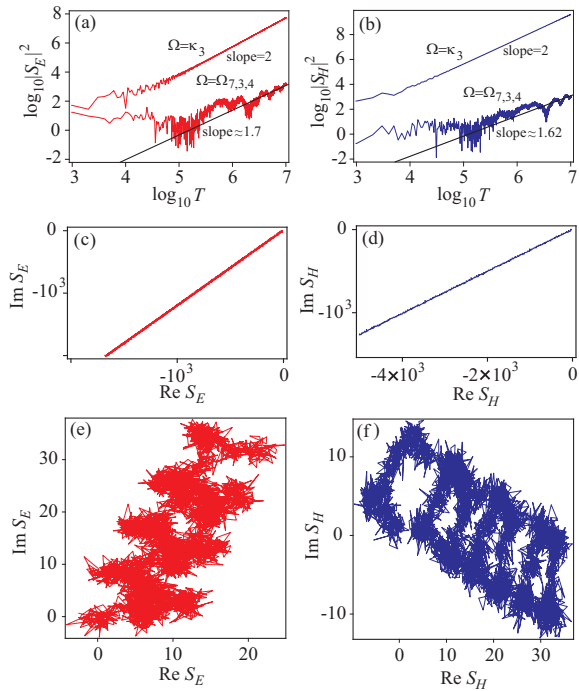}
\caption{(color online) Singular-continuous spectrum analysis of the
time series $\{\tilde{E}_{m}\}$ and $\{\tilde{H}_{m}\}$ for $\rho=1$.
(a)~$|S_{E}|^2$ and (b)~$|S_{H}|^2$ as functions of ${\rm log}_{10}T$ at $\Omega=\kappa_{3}$
and $\Omega=\Omega_{7,3,4}$. The paths of (c)~$S_{E}$ and (d)~$S_{H}$ in the complex planes
(${\rm Re}\,S_{E}$, ${\rm Im}\,S_{E}$) and
(${\rm Re}\,S_{H}$, ${\rm Im}\,S_{H}$), respectively, at $\Omega=\kappa_{3}$. The paths of (e)~$S_{E}$ and (f)~$S_{H}$
in the same planes at $\Omega=\Omega_{7,3,4}$.}
\end{figure}

\begin{figure}[h]
\includegraphics{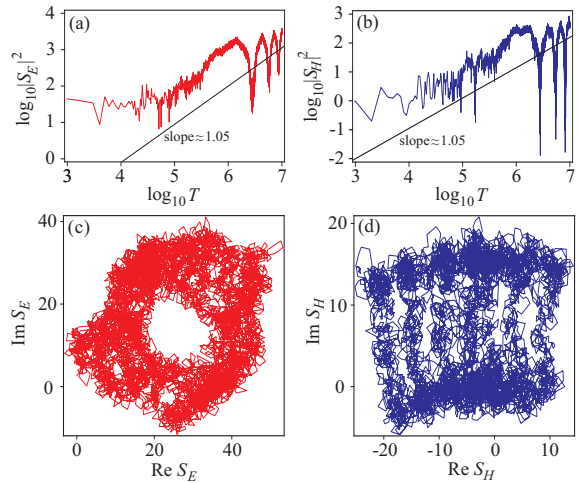}
\caption{(color online) Singular-continuous spectrum analysis of the
time series $\{\tilde{E}_{m}\}$ and $\{\tilde{H}_{m}\}$ for $\rho=0.5$.
(a)~$|S_{E}|^2$ and (b)~$|S_{H}|^2$ as functions of ${\rm log}_{10}T$ at
$\Omega=\Omega_{6,5,4}$.  The corresponding
paths of (c)~$S_{E}$ and (d)~$S_{H}$ in the complex planes.}
\end{figure}

We have found that at some frequencies, the spectrum has
the scaling exponents $\beta=\gamma=2$. Figures~2(a) and 2(b)
show $|S_{E}|^2$ and $|S_{H}|^2$ as functions of ${\rm log}_{10}T$ for
$\rho=1$ at one of such frequencies, namely, $\Omega=\kappa_{3}=8.65\ldots$~.
The corresponding paths in the complex planes are presented in
Figs.~2(c) and 2(d). Thus, in this case, we deal with a discrete
component of the spectrum. However, at the combination frequency
$\Omega_{7,3,4}=32.96\ldots$ (hereafter, $\Omega_{l,m,n}=l\Omega_{1}+
m\Omega_{2}+n\kappa_{2}$), we have $\beta\approx 1.7$ and $\gamma\approx 1.62$
[see Figs.~2(a) and 2(b)], and the behavior of the dependences
$|S_{E}|^2$ and $|S_{H}|^2$ in this case is typical of a singular
continuous component~\cite{P94,Yal,Agr}. The corresponding paths in Figs.~2(e) and
2(f) exhibit fractal structures. These results strongly
suggest that the considered spectrum of electromagnetic oscillations
is not purely discrete and contains singular continuous components.
For $\rho=1$, we have also found such components at, e.g., the combination
frequencies $\Omega_{3,3,4}$ ($\beta\approx 1.57$ and $\gamma=0$),
$\Omega_{3,4,4}$ ($\beta\approx 1{.}6$ and $\gamma=0$),
$\Omega_{6,3,4}$ ($\beta\approx 1{.}53$ and $\gamma\approx 1{.}4$),
$\Omega_{3,-1,4}$ ($\beta\approx 1{.}1$ and $\gamma\approx 1{.}55$),
$\Omega_{6,5,4}$ ($\beta\approx 1{.}52$ and $\gamma\approx 1{.}8$), and
$\Omega_{5,6,4}$ ($\beta\approx 1{.}4$ and $\gamma\approx 1{.}2$).
For many frequencies, a power-law growth of the spectrum is observed
with the exponents $\beta$ and $\gamma$ which differ from 2 only slightly.

We now pass to consideration of some spectral features of oscillations
inside a nonlinear resonator for $\rho=0.5$. Here, the components of
the singular continuous spectrum appear at higher frequencies
than for  $\rho=1$. The values of $\beta$ and $\gamma$ for $\rho=0.5$
turn out to be smaller than for $\rho=1$ at the same frequency.
For example, at  $\Omega=\Omega_{6,5,4}=35{.}02\ldots$,
we have $\beta\approx 1.52$ and $\gamma\approx 1.8$ for $\rho=1$, and
$\beta\approx 1{.}05$ and $\gamma\approx 1{.}05$ for $\rho=0.5$
[see Figs.~3(a) and 3(b)]. The corresponding curves in the complex
planes (${\rm Re}\,S_{E}$, ${\rm Im}\,S_{E}$) and
(${\rm Re}\,S_{H}$, ${\rm Im}\,S_{H}$), which are presented in Figs.~3(c) and 3(d),
display fractal behavior.

Thus, our analysis shows that the Fourier spectrum of the
electromagnetic oscillations in the cavity is a mixture of discrete
and singular continuous components. Similar phenomena have been reported in the
literature and discussed in, e.g., \cite{P94,Zak} as applied
to the dynamics described by forced maps and symbolic sequences.
In our case, calculations of the FFT and the autocorrelation
function (ACF) do not allow one to adequately investigate
the spectral properties of the considered oscillations.
Because of the weakness of the contribution from the singular
continuous part, the FFT and the ACF are similar to those of
almost periodic motion. Scaling is a powerful method to detect
the presence of a singular continuous spectrum. However, it is not
clear how this component can be separated from the discrete spectrum.

The existence of regimes with singular continuous (fractal) spectra
in dissipative dynamical systems described by discrete maps or ODEs
is well known~\cite{Din,P94,P95,Yal}. Such regimes corresponding
to strange nonchaotic attractors are realized on sets of positive
measure in the parameter spaces of dissipative dynamical systems and
are typical of the intermediate region between almost periodic
and random motions. Our study demonstrates that the nonlinear
dynamics with a singular continuous spectrum can occur in an
exactly integrable distributed nondissipative system.
We have found that the implicit functions given by Eqs.~(\ref{eq3})
and~(\ref{eq7}), which are exact solutions of system (\ref{eq1}),
are not almost periodic in $\tau$ and their Fourier spectra
contain singular continuous components. Studying these functions
is of great interest for physics and mathematics.
We have shown that such functions can be finite-amplitude
single-valued solutions of the Maxwell equations and, hence,
describe actually existing electromagnetic oscillations.
Thus, Eqs.~(\ref{eq3})
and~(\ref{eq7}) provide a new description of complex nonlinear
dynamics.

This work was supported by the Russian Foundation for Basic Research (Project No. 12--02--00904-a)
and the Ministry of Education and Science of the Russian Federation
(Contract Nos. P313 and 11.G34.31.0048). A.\,V.\,K. acknowledges partial support from the Greek Ministry of Education under the project THALIS (RF--EIGEN--SDR).

\bibliography{Petrov_Kudrin}

\end{document}